\documentclass[graybox, envcountchap]{svmult}

\usepackage{mathptmx}        % selects Times Roman as basic font
\usepackage{amsmath}
\usepackage{amssymb}
\usepackage{color}
\usepackage{helvet}          % selects Helvetica as sans-serif font
\usepackage{courier}         % selects Courier as typewriter font
\usepackage{dirtree}
\usepackage{makeidx}        % allows index generation
\usepackage{graphicx}        % standard LaTeX graphics tool
\usepackage{subfig}

\usepackage{multicol}        % used for the two-column index
\usepackage[bottom]{footmisc}% places footnotes at page bottom

\usepackage{hyperref}        %for hyperlinks
\hypersetup{colorlinks=true,urlcolor=blue}

\usepackage[misc]{ifsym}

\makeindex             % used for the subject index
\begin{document}

%%%%%%%%%%%%%%%%%%%%%%%%%%%%%%%%%%%%%%%%%%%%%%%%%%%%%%%%%%%%%%%%%

\title{Information retrieval from Euclidean path integral}

\author{Dong-han Yeom}

\institute{Department of Physics Education, Pusan National University, Busan 46241, Republic of Korea\\
Leung Center for Cosmology and Particle Astrophysics, National Taiwan University, Taipei 10617, Taiwan\\
\email{innocent.yeom@gmail.com}}

\maketitle

\abstract{In this article, we review the information loss paradox in the spirit of the Euclidean path integral approach. First, we argue that there is a long debate about the information loss paradox, and the non-perturbative quantum gravitational wave function must include the clue to the paradox. The Euclidean path integral approach provides the best way to describe the wave function. From this wave function, we can notice that there are not only semi-classical but also non-perturbative contributions, which are highly suppressed but preserved information. Information retrieval will be sufficiently explained if such non-perturbative contributions must be dominated by the late time. We will show that there is sufficient evidence that this scenario can be realized in generic circumstances. Finally, we compare this scenario with alternative approaches. Also, we comment on some unresolved issues that need to be clarified.}

%%%%%%%%%%%%%%%%%%%%%%%%%%%%%%%%%%%%%%%%%%%%%%%%%%%%%%%%%

\section{Introduction}
\label{sec:1}

The information loss paradox \cite{Hawking:1976ra} is one of the fundamental questions that must be resolved by the quantum theory of gravity, or the Theory of Everything. Black holes will evaporate due to Hawking radiation in a finite time \cite{Hawking:1975vcx}. However, Hawking radiation seems to depend only on the information of its horizon, e.g., the mass $M$, the charge $Q$, and the angular momentum $J$. After the evaporation, can Hawking radiation carry quantum mechanical information? If not, we will eventually lose the unitarity of quantum mechanics and the fundamental predictability. On the other hand, if information must be preserved, how can Hawking radiation carry information without causing further inconsistency? This paradox drastically reveals the tension between quantum mechanics and general relativity. By struggling to reconcile this tension, we will learn many properties that must be included in the consistent quantum theory of gravity.

This article will use the Euclidean path integral approach \cite{Hartle:1983ai} to study the information loss problem. We will summarize the information loss paradox before discussing further details of this approach. We will briefly discuss why the information loss paradox is difficult to solve and what the logically possible candidate answers are.

\subsection{Information loss paradox}

We define the information loss paradox as an inconsistency between well-confirmed theories. We may summarize such well-confirmed theories or assumptions as follows \cite{Chen:2014jwq}.
\begin{itemize}
\item[--] 1. The entire process of black hole evaporation satisfies \textit{unitarity}.
\item[--] 2. Spacetime satisfies equations according to \textit{general relativity} except for a singularity. 
\item[--] 3. \textit{Local quantum field theory} is satisfied near the horizon, and Hawking radiation is emitted from the horizon.
\item[--] 4. The statistical entropy of a black hole is proportional to the area of the black hole $A$, i.e., $\log \Omega = A/4$, where $\Omega$ is the number of accessible states of the black hole.
\item[--] 5. There is an observer who can collect, count, and compute information or entropy from Hawking radiation.
\end{itemize}

These five assumptions cause inconsistency. Let us think that a black hole is formed due to a gravitational collapse. The collapsed matter will carry some quantum information. Due to Assumptions 1 and 3, a black hole should emit Hawking radiation, and the Hawking radiation should have information about collapsed matter after a specific time. On the other hand, according to Assumption 2 and Assumption 3, the collapsed matter will satisfy the local quantum field theory and general relativity; hence, without experiencing a dramatic event, information will be safely carried to the singularity.

Then, it seems there are two copies of information, one inside the horizon and the other outside. According to Assumptions 1 and 3, the outside should have information; according to Assumptions 2 and 3, the inside should have the original information, too. This violates the no-cloning theorem of quantum mechanics.

However, this naive argument might have some loopholes. When will the black hole emit information? Can any witness see the copied information if it is duplicated, violating the no-cloning theorem? We need to discuss the entanglement entropy and the Page curve to answer these questions.

\begin{figure}[b]
\sidecaption
\includegraphics[scale=.6]{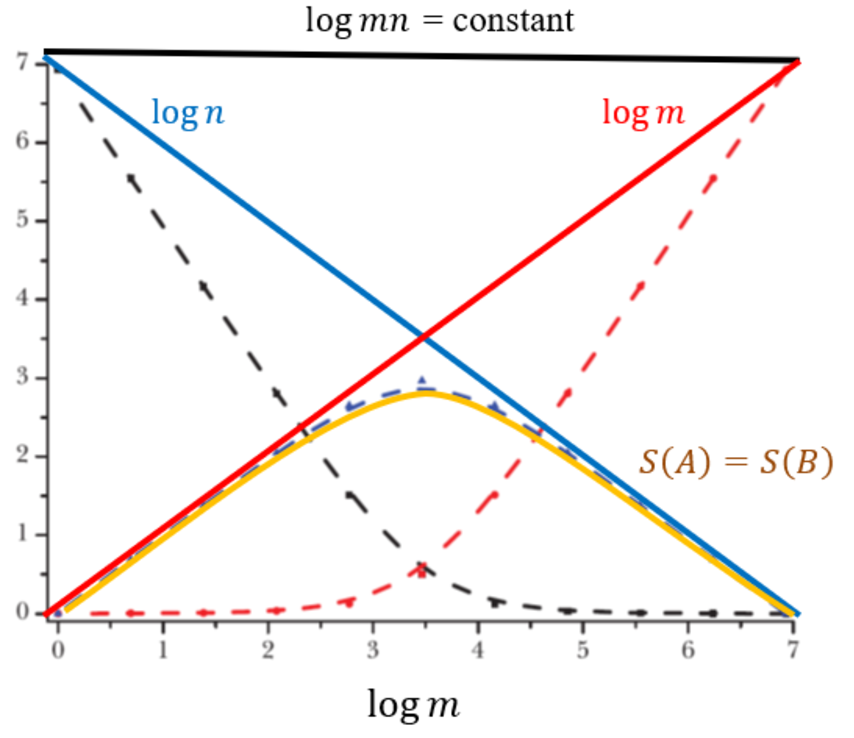}
\caption{Boltzmann entropies of radiation ($\log n$), black hole ($\log m$), and the entanglement entropy ($S(A) = S(B)$). The total number of states is a constant ($\log mn$). The red and black dashed curves are $I_{A}$ and $I_{B}$, respectively.}
\label{fig:fig1}
\end{figure}

\subsection{Entanglement entropy and the Page time}

Let us ask how to quantify the amount of information that is emitted during the evaporation process. To answer this question, let us first define a bipartite system, where outside the black hole is $A$ and inside the black hole is $B$. One can determine the total density matrix as $\rho$. From this, one can define the density matrix of a subsystem $\rho_{A} = \mathrm{tr}_{B}~\rho$ and $\rho_{B} = \mathrm{tr}_{A}~\rho$. From this, one can define the entanglement entropy between $A$ and $B$:
\begin{eqnarray}
S_{A} = - \mathrm{tr}_{A} \rho_{A} \log \rho_{A},
\end{eqnarray}
where $S_{A} = S_{B}$ and $S_{AB} = 0$ for a pure state. The maximum bound of the entanglement entropy is the Boltzmann entropy of the system, i.e., if the number of states of $A$, $B$ is $m$, $n$, respectively, then
\begin{eqnarray}
S_{A} &\leq& \log m,\\
S_{B} &\leq& \log n
\end{eqnarray}
are satisfied.

There is no information if the radiation is thermal, and the entanglement entropy is the same as its Boltzmann entropy. On the other hand, if there is a deviation between the Boltzmann entropy and the entanglement entropy, it is possible to distinguish information in principle. Hence, we can define the \textit{thermodynamic depth} \cite{Lloyd:1988cn}, or information of subsystem, as follows:
\begin{eqnarray}
I_{A} \equiv \log m - S_{A}.
\end{eqnarray}
Of course, $I_{A}$ is positive definite and is zero only if the state is perfectly thermal (Fig.~\ref{fig:fig1}).

As the black hole evaporates, one can think about the three kinds of information: (1) information of part $A$, i.e., $I_{A}$, (2) information of part $B$, i.e., $I_{B}$, and (3) information of entanglement, i.e., the mutual information $I_{A:B}$, where
\begin{eqnarray}
I_{A:B} \equiv S_{A} + S_{B} - S_{AB}.
\end{eqnarray}
For an information-preserving system or a pure state, the sum of three pieces of information will be conserved \cite{Hwang:2016otg}:
\begin{eqnarray}
I_{A} + I_{B} + I_{A:B} = \log mn,
\end{eqnarray}
where $mn = \mathrm{const.}$ in a pure state.

Now, we can ask about the relationship between entanglement entropy and Boltzmann entropy. Of course, this depends on the details of its quantum state, but assuming the most typical or random state, one can estimate typical relations. According to Page \cite{Page:1993df}, if $m < n$, approximately,
\begin{eqnarray}
S_{A} \simeq \log m,
\end{eqnarray}
where analytic and numerical methods can easily confirm this result. This implies that if $m < n$, then $I_{A} \simeq 0$; hence, Hawking radiation does not carry sufficient information. After the black hole emits more than half of the number of states, in other words, if $m > n$, then $I_{B} \simeq 0$ and $I_{A}$ should grow linearly and eventually all information must be escaped. Hence, it is reasonable to explain that (1) in the early stage of evaporation, almost all information is inside the horizon ($I_{B}$ is dominated), (2) as time goes on, inside and outside of the black hole are entangled and the mutual information is dominated ($I_{A:B}$ is dominated), and (3) eventually, as the number of states of outside dominates, all information will escape from inside to outside by Hawking radiation ($I_{A}$ is dominated).

This is quite a reasonable explanation of information transfer from inside to outside the black hole. However, naturally, one can raise the following question: \textit{when can we distinguish information from Hawking radiation}? Up to now, there is no principle to answer about this. However, if we introduce Assumption 4, i.e., if the Boltzmann entropy is proportional to its areal entropy, one can estimate the time scale, so-called the \textit{Page time} \cite{Page:1993wv}:
\begin{eqnarray}
\tau_{P} \simeq M^{3}.
\end{eqnarray}
This time scale is long, but the black hole can still be semi-classical. Therefore, Hawking radiation is the only channel that carries information around the Page time.

Therefore, based on Assumptions 1 - 4, we can conclude that the black hole should emit information when the black hole is still in the semi-classical scale. In the following subsection, let us study why this result will cause inconsistency.

\subsection{Trouble with entanglements}

We can see the details of inconsistency between assumptions in several ways. We clarify this issue in the following three ways. These three approaches are slightly different from each other but present the same aspect of the paradox of black hole information.

\begin{figure}[b]
\sidecaption
\includegraphics[scale=.5]{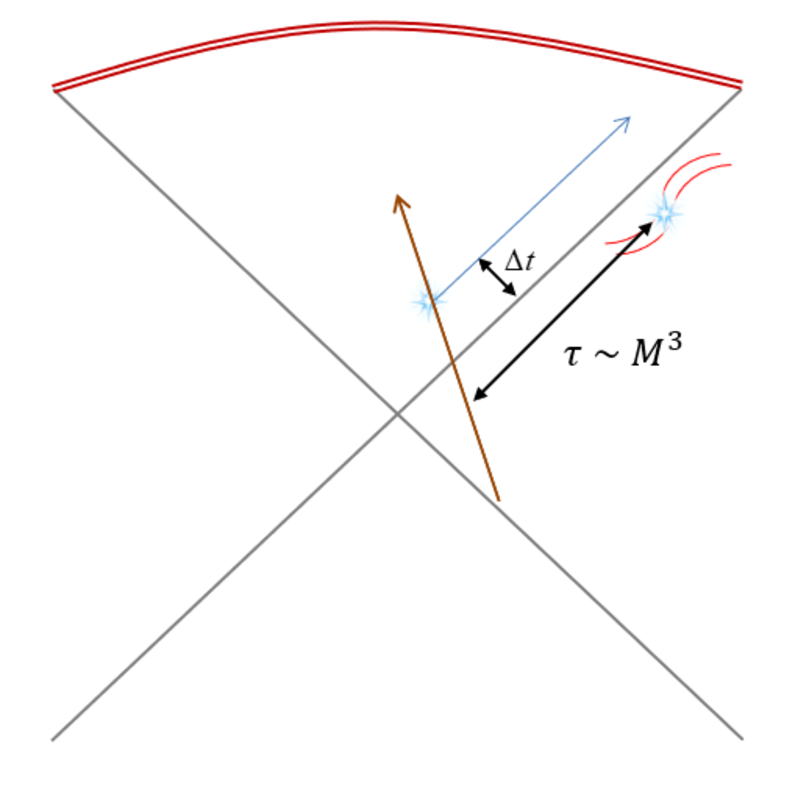}
\caption{Duplication experiment: after a bit of information falls into the horizon, it escaped after the Page time $\tau$. An observer can measure the duplicated information if the original signal is sent to the outgoing direction inside the horizon between the time $\Delta t$.}
\label{fig:fig2}
\end{figure}

\begin{itemize}
\item[--] 1. \textit{Duplication of information} \cite{Yeom:2009zp}: According to Assumption 5, there can be an observer who can measure information from Hawking radiation. Let us assume that an asymptotic observer measures information from Hawking radiation after the Page time. After the observer measures information, let us assume that the observer falls into the black hole. The question is whether the original information that collapsed into the black hole can be sent to the observer. If it is, in principle, allowed, the observer can witness the violation of the unitarity, which reveals the inconsistency among the assumptions.

The possibility of this \textit{duplication experiment} is related to the uncertainty principle \cite{Susskind:1993mu}. To successfully send information to the observer, the collapsed matter should send a signal after the time scale $\Delta t$ (Fig.~\ref{fig:fig2}), where
\begin{eqnarray}
\Delta t \simeq e^{-\frac{\tau}{r}},
\end{eqnarray}
where $r \simeq M$ is the spatial size of the black hole and $\tau$ is the time-scale that emits information, e.g., $\tau_{P} \sim M^{3}$, or the temporal size of the black hole. Quantum mechanically, to send some information during the time-scale $\Delta t$, one needs energy $\Delta E \sim 1 / \Delta t$ due to the uncertainty principle. If $\Delta E$ is too large, so to speak, if $\Delta E > M$, then this duplication experiment is, in principle, disallowed.

However, in the semi-classical regime, this is not prevented in principle \cite{Yeom:2008qw,Hong:2008mw,Yeom:2009zp}. Consider a number $N$ of matter fields contributing to Hawking radiation. In addition, let us scale the spatial size of the black hole such as $r \rightarrow r' = \sqrt{N} r$. Then, the temporal scale will be stretched by $\tau \sim M^{3} \rightarrow \tau' = (\sqrt{N}M)^{3}/N = \sqrt{N} M^{3} \sim \sqrt{N} \tau$. This implies that, in the large $N$ limit, the following quantity is invariant:
\begin{eqnarray}
\frac{\tau}{r} = \frac{\tau'}{r'},
\end{eqnarray}
while the spatial and temporal size of the black hole increases in the Planck units (hence, semi-classicality is still satisfied). Therefore, $\Delta t \rightarrow \sqrt{N} \Delta t$ and $\Delta E \rightarrow 1 / (\sqrt{N} \Delta t)$ is obtained. By choosing a parametrically large $N$, one can obtain the condition that $\Delta E < M$. If we think about the time-scale of information emission using the scrambling time \cite{Hayden:2007cs}, $\tau_{S} \sim M \log M$, then even with a small number of matter fields $N$, this energy condition $\Delta E < M$ might be easily satisfied, and hence, the duplication experiment is possible \cite{Yeom:2009zp}. (Also, it is possible to construct such a successful duplication experiment without assuming extremely large $N$ in lower dimensional examples, e.g., see \cite{Kim:2013fv}.) Therefore, it is evident that the five assumptions will eventually be inconsistent.

\begin{figure}[t]
\sidecaption
\includegraphics[scale=.6]{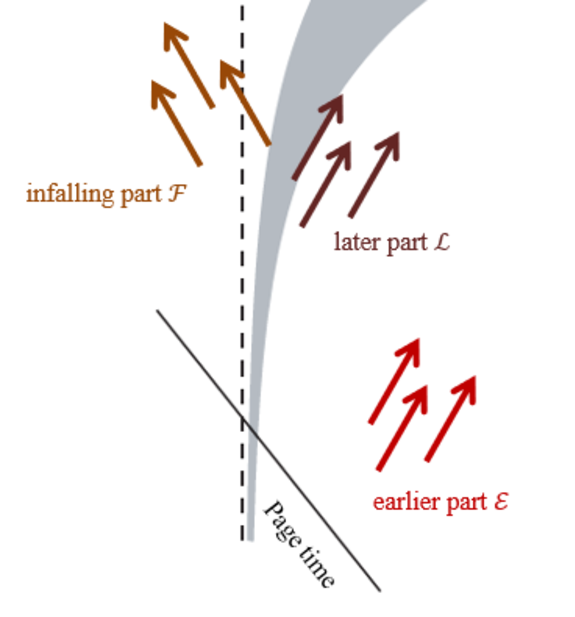}
\caption{Three parts for the entanglement entropy: radiation before the Page time $\mathcal{E}$, after the Page time $\mathcal{L}$, and its counterpart $\mathcal{F}$.}
\label{fig:fig3}
\end{figure}

\item[--] 2. \textit{Inconsistency of entanglement entropy relations} \cite{Almheiri:2012rt}: Let us distinguish the three systems: Hawking radiation before the Page time $\mathcal{E}$, Hawking radiation after the Page time $\mathcal{L}$, and its counterpart $\mathcal{F}$ (Fig.~\ref{fig:fig3}). For these three systems, the strong subadditivity and subadditivity relations must be satisfied:
\begin{eqnarray}
S_{\mathcal{L}} + S_{\mathcal{ELF}} &\leq& S_{\mathcal{EL}} + S_{\mathcal{LF}},\\
\left| S_{\mathcal{E}} - S_{\mathcal{LF}} \right| &\leq& S_{\mathcal{ELF}} \leq S_{\mathcal{E}} + S_{\mathcal{LF}}.
\end{eqnarray}

First, due to Assumptions 1, 3, and 4, after the Page time, information should be emitted. This means that the entanglement entropy should begin to decrease. Note that $\mathcal{E} \cup \mathcal{L}$ is the entire Hawking radiation, and hence $S_{\mathcal{EL}}$ is the entanglement entropy of Hawking radiation. This should decrease from its maximum value $S_{\mathcal{E}}$, i.e., the entanglement entropy at the Page time. Hence, we obtain $S_{\mathcal{EL}} < S_{\mathcal{E}}$.

Second, due to Assumptions 1, 2, and 3, Hawking radiation does not violate the unitarity. Hence, the only possible way to create particle and anti-particle pairs is that the particle ($\mathcal{L}$) and anti-particle ($\mathcal{F}$) pairs are separable states, i.e., the created particles are not entangled with the background system. Therefore, $S_{\mathcal{LF}} = 0$ is required.

Then, by applying the strong subadditivity and subadditivity relations, one can conclude that $S_{\mathcal{L}} < 0$; of course, this result is inconsistent. According to Assumption 5, there is an observer who measures entanglement entropy of $\mathcal{L}$ and the observer will notice that some of the previous assumptions must be wrong because $S_{\mathcal{L}}$ cannot be negative definite. (See also an analog problem in moving mirror cases, in \cite{Chen:2017lum}.)

\begin{figure}[b]
\sidecaption
\includegraphics[scale=.6]{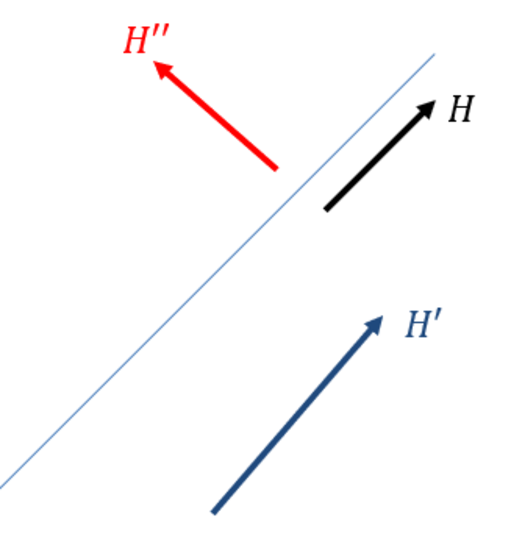}
\caption{A Hawking particle $H$ that emitted after the Page time is maximally entangled with an earlier radiation $H'$, and at the same time, maximally entangled with its counterpart $H''$.}
\label{fig:fig4}
\end{figure}

\item[--] 3. \textit{Monogamy of entanglements} \cite{Maldacena:2013xja}: According to Assumptions 1, 3, and 4, after the Page time, the entanglement entropy must decrease as time goes on. This means that, after the Page time, the number of states outside the horizon should increase monotonically (due to continuous evaporation) while the entanglement entropy of Hawking radiation decreases. The only logically possible way is for there to be entanglements between emitted particles. Therefore, after the entanglement purification protocol, one can find a maximally entangled pair (say, $H$ and $H'$) among Hawking particles after the Page time.

Assumptions 1, 2, and 3 show that a Hawking radiation pair (particle and anti-particle pair) cannot violate the unitarity. Hence, the only possible way to create them is for the pair to form a separable state. Therefore, the emitted Hawking particle (say, $H$) should be maximally entangled with its counterpart (say, $H''$).

However, this causes an inconsistency. After the Page time, the emitted Hawking particle $H$ is maximally entangled with its earlier part $H'$; simultaneously, the emitted Hawking particle $H$ is maximally entangled with its counterpart $H''$. However, this is disallowed due to the monogamy of quantum entanglements. According to Assumption 5, if an observer detects $H'$ and finds out $H''$ inside the horizon, the observer will witness the violation of assumptions. This monogamy argument is deeply related to the previous entanglement entropy argument \cite{Hwang:2017yxp}.
\end{itemize}

These three arguments agree that five assumptions cause problems in terms of information or entanglements. Which assumption should be modified?

\subsection{Candidate resolutions}

Logically, one can classify the candidate resolutions of the paradox as five possibilities \cite{Ong:2016iwi}.
\begin{itemize}
\item[--] 1. \textit{Unitarity is violated} \cite{Unruh:1995gn}. However, this requires further explanations. In addition, the violation of the unitarity principle is tense with the AdS/CFT correspondence \cite{Maldacena:1997re}.
\item[--] 2. \textit{General relativity is violated at a considerable length scale.} For example, a firewall \cite{Almheiri:2012rt}, a fuzzball \cite{Mathur:2005zp}, a stretched horizon \cite{Susskind:1993if}, a membrane \cite{Callan:1996dv}, a large remnant \cite{Chen:2014jwq}, or any other strange object appears at the horizon scale. However, these objects can affect future infinity and, perhaps, can be falsified from astrophysical experiments \cite{Hwang:2012nn,Chen:2015gux}. 
\item[--] 3. \textit{Quantum effects can be non-local} \cite{Page:2013mqa}. In this case, one must explain in which circumstances such a non-local effect is triggered, even within the semi-classical regime.
\item[--] 4. \textit{The areal entropy does not represent the Boltzmann entropy} \cite{Chen:2014jwq}. This possibility may result in the remnant picture. However, one can ask whether the remnant can carry sufficiently large entropy; if the remnant is the Planck scale, then it might be infinitely produced; also, one can ask whether the remnant scenario can successively work in generic situations \cite{Chen:2014jwq}. 
\item[--] 5. \textit{No observer can detect information} \cite{Sasaki:2014spa,Chen:2015lbp}. If this is the case, then one needs to ask how to realize this possibility and whether such a realization mechanism can avoid the inconsistency arguments of the previous subsection.
\end{itemize}

In this article, we may focus on the possibility that these five possibilities are \textit{all true in a way}. What does it mean?

\subsection{The answer is in the wave function}

The previous inconsistency arguments reveal the trouble between \textit{semi-classical black hole dynamics} and the \textit{unitarity of quantum mechanics}.

Let us analogously see the particle and wave nature of quantum mechanics. Of course, two properties are inconsistent, but in reality, there is no contradiction between the two natures because there is a \textit{quantum measurement} that distinguishes two sides; before the measurement, we only see the wave nature, while after the measurement, we only see the particle nature. The unitarity is related to the wave nature, while classical dynamics is associated with the particle nature.

Can we say that the same thing happens in the black hole physics? This requires the \textit{quantum gravitational wave function} that explains the entire black hole dynamics \cite{Chen:2022eim}. In terms of the wave function, losing information is impossible; hence, the unitarity must be preserved. However, in this wave function picture, we must sum over all classical contours/histories; therefore, the semi-classical equations cannot be satisfied. Therefore, Assumption 1 is satisfied, while Assumptions 2 and 3 are no more apparent; hence, it is not surprising to expect non-local physics or dramatic effects at the horizon scale because semi-classical equations are no longer valid.

On the other hand, what `we' will measure or observe is a semi-classical history. Of course, this semi-classical history should satisfy Assumptions 2 and 3, i.e., general relativity and local quantum field theory. However, it is not surprising that this semi-classical history may not fulfill the unitarity because we do not see the entire wave function of the universe.

If this picture is correct, then there are two points of view: the wave nature and the semi-classical nature. Hence, there are two kinds of observers: the \textit{unitary observer} who can see the entire wave function and the \textit{semi-classical observer} who can follow only one semi-classical history. We are semi-classical observers, and hence, we will lose information. Still, the global wave function will carry information anyway, and the unitarity must be satisfied eventually, although it is beyond our empirical scopes.

This scenario seems to provide an exciting point of view that smoothly avoids the previous tensions. A big question remains: how can we consistently compute the entire wave function? Of course, at this moment, we do not have a consistent theory of everything, and it is fair to say that we do not know how to deal with this problem. However, there is an excellent way to see the brief structure of the wave function, at least approximately. The highway to glimpse the quantum gravitational wave function is to follow the \textit{Euclidean path integral approach}. Now, let us start to discuss what it looks like.

\newpage

\section{The wave function of the Universe}
\label{sec:2}

There are two main approaches to quantum gravity. First, one considers perturbations around a fixed classical background and computes expectation values of perturbations using Feynman diagrams \cite{DeWitt:1967ub}. Second, one directly solves the wave function by introducing the quantum Hamiltonian constraint equation \cite{DeWitt:1967yk}. The former is called by the \textit{covariant approach} because one can use the path integral and the path integral has the manifestly covariant Lagrangian; the latter is called by the \textit{canonical approach}, because this requires the Hamiltonian and canonical quantization relations, etc. Both methods have their problems, e.g., the non-renormalizability problem, and at the same time, some efforts are being made to avoid the issues, e.g., the string theory or loop quantum gravity.

To see the entire dynamics of black hole physics, we need the non-perturbative nature of the wave function, and hence, we should rely on the canonical approach. According to the canonical approach of quantum gravity, for a given 3-hypersurface $h_{ab}$ and field values at the hypersurface $\phi$, every information is in the wave function $\Psi[h_{ab}, \phi]$, where this should satisfy
\begin{eqnarray}
\hat{\mathcal{H}} \Psi[h_{ab},\phi] = 0,
\end{eqnarray}
where $\hat{\mathcal{H}}$ is the quantum Hamiltonian constraint operator; this equation is known as the \textit{Wheeler-DeWitt equation} \cite{DeWitt:1967yk}. There are also quantum momentum constraint equations; sometimes, these are very important to understand the effects of quantum gravity. However, this article will focus on the Wheeler-DeWitt equation to cover the system's dynamics.

This Wheeler-DeWitt equation is indeed very complicated and perhaps ill-defined because this is not a simple differential equation but a \textit{functional} differential equation. Avoiding this nature of functional differential equations is impossible except for cosmological cases \cite{Kuchar:1994zk}. Then, how can we solve this wave function? We do not know the best answer yet, but writing a \textit{formal} solution is possible. That is to use the \textit{path integral}.

\subsection{Euclidean path integral approach}

In some sense, the Euclidean path integral \cite{Hartle:1983ai} solves the equation of the canonical approach non-perturbatively, using the path integral that is manifestly covariant; hence, the Euclidean path integral approach is between two approaches.

One can write the propagator between an in-state $| h_{ab}^{\mathrm{in}}, \phi^{\mathrm{in}} \rangle$ to an out-state $| h_{ab}^{\mathrm{out}}, \phi^{\mathrm{out}} \rangle$
\begin{eqnarray}
\left\langle h_{ab}^{\mathrm{out}}, \phi^{\mathrm{out}} | h_{ab}^{\mathrm{in}}, \phi^{\mathrm{in}} \right\rangle = \Psi \left[ h_{ab}^{\mathrm{out}},\phi^{\mathrm{out}}; h_{ab}^{\mathrm{in}}, \phi^{\mathrm{in}} \right] = \int \mathcal{D}g_{ab} \mathcal{D}\Phi e^{iS[g_{\mu\nu},\Phi]},
\end{eqnarray}
where $S[g_{\mu\nu},\Phi]$ is the action with the metric $g_{\mu\nu}$ and a matter field $\Phi$, and this path integral is the sum over all histories that connect from the in-state $| h_{ab}^{\mathrm{in}}, \phi^{\mathrm{in}} \rangle$ to the out-state $| h_{ab}^{\mathrm{out}}, \phi^{\mathrm{out}} \rangle$. This is a path integral and, hence, is a formal solution of the Wheeler-DeWitt equation.

However, one problem is that this path integral is badly behaved. A good strategy to compute the wave function is to Wick-rotate the time from Lorentzian to Euclidean, $dt = - id\tau$, and to obtain
\begin{eqnarray}
\left\langle h_{ab}^{\mathrm{out}}, \phi^{\mathrm{out}} | h_{ab}^{\mathrm{in}}, \phi^{\mathrm{in}} \right\rangle = \int \mathcal{D}g_{ab} \mathcal{D}\Phi e^{- S_{\mathrm{E}}[g_{\mu\nu},\Phi]}.
\end{eqnarray}
In many examples of quantum mechanics, this Wick-rotation explains the ground state wave function \cite{Hartle:1983ai}. In the quantum gravitational case, it is impossible to define the ground state. However, we can analogously regard that the Euclidean path integral describes a putative ground state or a state with thermal equilibrium \cite{Gibbons:1976ue}.

The exact computation of the Euclidean path integral is still complicated, but now it is easier to \textit{approximate} the wave function. We can introduce the steepest-descent approximation by plugging on-shell histories, i.e., by summing over \textit{instantons} \cite{Hartle:1983ai}:
\begin{eqnarray}
\left\langle h_{ab}^{\mathrm{out}}, \phi^{\mathrm{out}} | h_{ab}^{\mathrm{in}}, \phi^{\mathrm{in}} \right\rangle \simeq \sum_{\mathrm{on-shell}} e^{- S_{\mathrm{E}}^{\mathrm{on-shell}}[g_{\mu\nu},\Phi]}.
\end{eqnarray}
These instantons can show the wave function's skeleton structures, including both perturbative and non-perturbative effects.

\subsection{Hawking radiation as instantons}

\begin{figure}[t]
\sidecaption
\includegraphics[scale=.45]{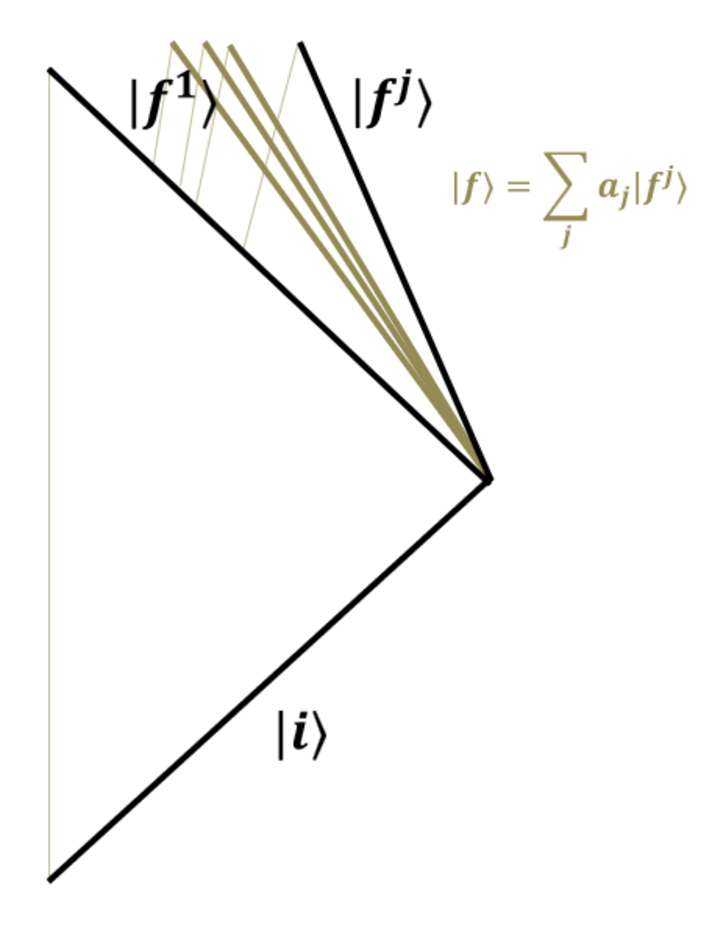}
\caption{The Euclidean path integral connects from the in-state $| i \rangle$ to the out-state $\left| f \right\rangle = \sum_{j} a_{j} \left| f^{j} \right\rangle$, where the out-state is a superposition of several classical boundaries.}
\label{fig:fig5}
\end{figure}

Let us apply this Euclidean path integral approach to black hole dynamics. We should start from an initial state that describes gravitational collapses, but for simplicity, we assume that the classical collapsing process was finished; hence, let us start with the initial state of the Schwarzschild black hole with mass $M$:
\begin{eqnarray}
\left| i \right\rangle = (\mathrm{Schwarzschild~black~hole~with~mass~} M).
\end{eqnarray}
It will evolve to an out-state $| f \rangle$ (supposed to be defined at the future infinity), where this out-state is in general a superposition of classical boundaries $| f^{j} \rangle$ ($j= 1, 2, ...$):
\begin{eqnarray}
\left| f \right\rangle = \sum_{j} a_{j} \left| f^{j} \right\rangle,
\end{eqnarray}
where $p_{j} \equiv |a_{j}|^{2}$ is the probability to transit from $| i \rangle$ to $| f^{j} \rangle$. Now, from the Euclidean path integral approach, one can write the relation:
\begin{eqnarray}
a_{j}^{*} = \left\langle f^{j} | i \right\rangle \simeq \sum_{| i \rangle \rightarrow | f^{j} \rangle} e^{- S_{\mathrm{E}}^{\mathrm{on-shell}}},
\end{eqnarray}
where we sum over all instantons that connects from $| i \rangle$ to $| f^{j} \rangle$ (Fig.~\ref{fig:fig5}).

Now let us define $| f^{j} \rangle$ as follows:
\begin{eqnarray}
\left| f^{j} \right\rangle &=& (\mathrm{Schwarzschild~black~hole~with~mass~} M') \nonumber \\
 &&+ (\mathrm{matter~field~emission~with~energy~}\delta M),
\end{eqnarray}
where $M - M' = \delta M > 0$. Then, how can we compute $a_{j}$?

To compute the transition probability, what we need to consider are as follows \cite{Chen:2018aij}:
\begin{itemize}
\item[--] 1. \textit{Action}: The black hole can emit a scalar field, a photon field, a graviton field, or other massive particles, etc. To consider the simplest case, let us assume that the action contains a free scalar field:
\begin{eqnarray}
S = \int_{\mathcal{M}} \sqrt{-g} dx^{4} \left[ \frac{R}{16\pi} - \frac{1}{2} \left(\nabla \Phi\right)^{2} \right] + \int_{\partial \mathcal{M}} \frac{K - K_{0}}{8\pi} \sqrt{-h} dx^{3},
\end{eqnarray}
where $\mathcal{M}$ is the 4-manifold, $\partial\mathcal{M}$ is the boundary of $\mathcal{M}$, $R$ is the Ricci scalar, and $K$ and $K_{0}$ are the Gibbons-Hawking boundary terms of the solution and the Minkowski background (evaluated only at infinity) \cite{Gibbons:1976ue}, respectively.

If we consider a Euclidean on-shell solution with a free scalar field, the four-volume integration term automatically vanishes, and we obtain
\begin{eqnarray}
S = \int_{r = \infty} \frac{K - K_{0}}{8\pi} \sqrt{-h} dx^{3} + (\mathrm{boundary~term~at~horizon}),
\end{eqnarray}
where the first term is evaluated at infinity.

\item[--] 2. \textit{Complexification}: At the future infinity, we will consider an outgoing pulse of the matter field. If we evolve the solution backward in time, what will happen? We need to find out about Euclidean geometries. There might be many examples, but one of the trivial examples is that one Wick-rotates to the Euclidean Schwarzschild black hole with mass $M'$. Then, the matter field should Wick-rotate to the Euclidean time on the background of the Euclidean Schwarzschild black hole with $M'$ \cite{Chen:2018aij} (Fig.~\ref{fig:fig6}).

\begin{figure}[t]
\sidecaption
\includegraphics[scale=.45]{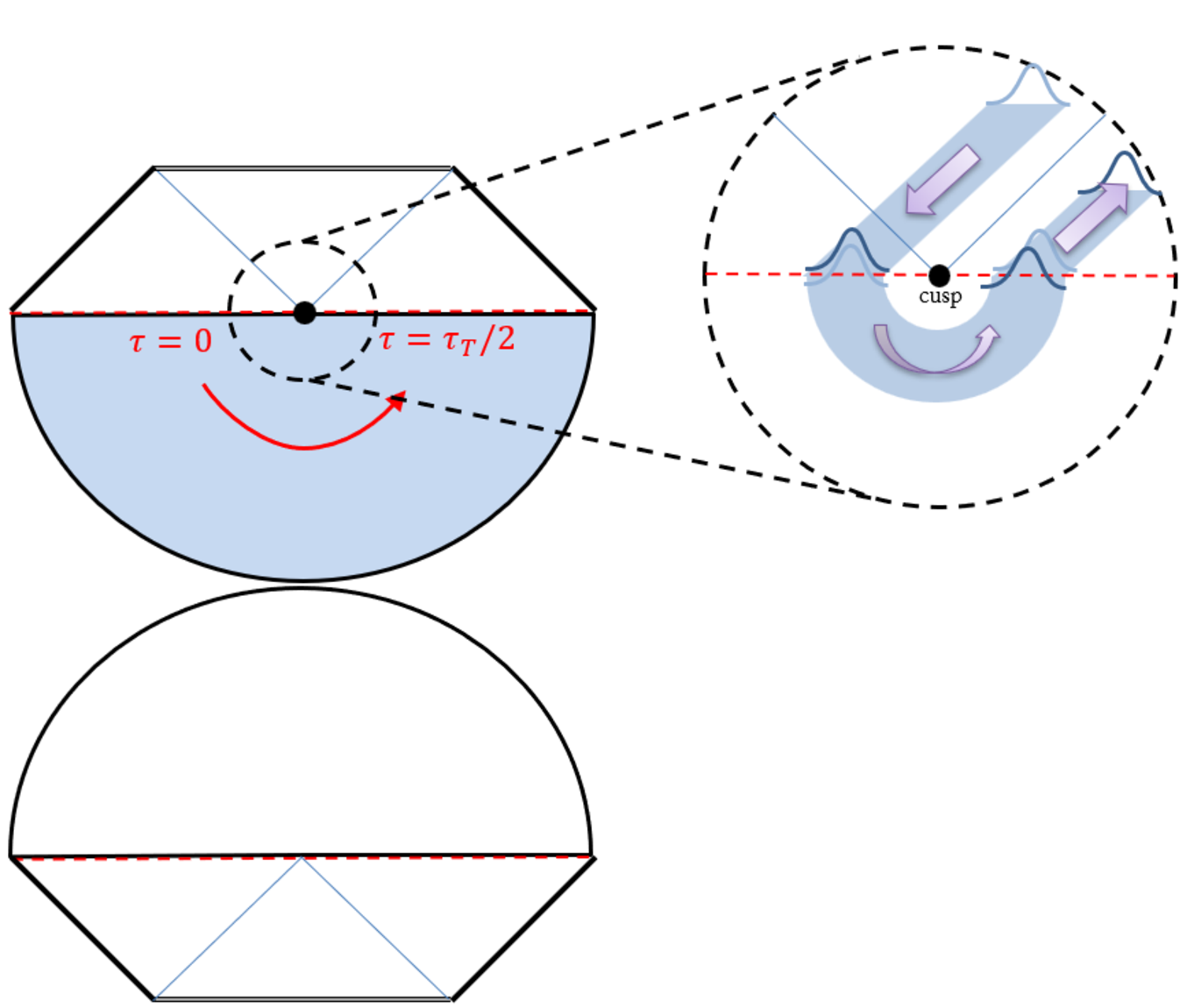}
\caption{The steepest-descent approximation that connects from a black hole with mass $M$ to a black hole with mass $M'$ and radiation $\delta M$. At a constant time hypersurface (red dashed line), we inserted Euclidean Schwarzschild spacetime for past and future diagrams (with the Euclidean time period $\tau_{T}$). There must be a cusp at the horizon (thick black dot). The scalar field will propagate along the arrows in the black dashed circle. Imposing the classicality condition outside the horizon, the scalar field must be complexified at the Euclidean domain and inside the horizon.}
\label{fig:fig6}
\end{figure}

The problem is that the matter field usually has non-vanishing differentials for time at the Wick-rotation point. Therefore, due to the Cauchy-Riemann or analyticity conditions, the matter field must be complex-valued after the Wick-rotation; hence, the metric must also be complexified at the Euclidean domain.

This might be strange, but this is already expected in the Euclidean path integral approach, and indeed, it is an essential property for a consistent Euclidean path integral approach \cite{Hartle:2008ng}. In the path integral, we will observe the in-state and out-state; in the intermediate process, we do not know what happens. So, it is unsurprising that some strange things happen in the intermediating process. However, we must require the reality condition at infinity \cite{Yeom:2021twr}; this condition is sometimes called by the \textit{classicality condition} \cite{Hartle:2008ng}, which must be needed for the future infinity. Note that this complexification does not change the computation of Euclidean actions.

\item[--] 3. \textit{Regularization of cusps}: Now the final task is to compute the Euclidean action from a Schwarzschild black hole with mass $M$ to a Schwarzschild black hole with mass $M'$ with matter emission $\delta M$. Since the Euclidean Schwarzschild black holes with mass $M$ and mass $M'$ are disconnected in Euclidean signatures, one can compute the transition probability as
\begin{eqnarray}
p_{j} = \left|\left\langle f^{j} | i \right\rangle\right|^{2} \simeq e^{- 2 \left( S_{\mathrm{E}}^{f^{j}} - S_{\mathrm{E}}^{i} \right)},
\end{eqnarray}
where $S_{\mathrm{E}}^{f^{j}}$ and $S_{\mathrm{E}}^{i}$ are the Euclidean action of $|f^{j}\rangle$ and $|i\rangle$, respectively.

As we have mentioned, the bulk integration does not contribute to the Euclidean action, but there are contributions at the boundaries of the manifolds. For $S_{\mathrm{E}}^{i}$, this is an Euclidean Schwarzschild manifold, and hence, the only contribution is at infinity (the horizon is regular). On the other hand, for $S_{\mathrm{E}}^{f^{j}}$, the infinity is indistinguishable from $S_{\mathrm{E}}^{i}$ because the asymptotic mass is the same, but at the horizon, the local mass is $M'$. This causes the mismatch of the Euclidean time period; asymptotically, the Euclidean time period is $8\pi M$, while the regularity condition required at the horizon is $8\pi M'$ \cite{Chen:2017suz}.

Due to this mismatch, we require a regularization technique at the horizon \cite{Gregory:2013hja}. As a result, when we evaluate $S_{\mathrm{E}}^{f^{j}} - S_{\mathrm{E}}^{i}$, the boundary term at infinity will be canceled perfectly and the boundary term at the horizon (in $S_{\mathrm{E}}^{f^{j}}$) will contribute only. After the regularization, we obtain that
\begin{eqnarray}
2 \left( S_{\mathrm{E}}^{f^{j}} - S_{\mathrm{E}}^{i} \right) = \frac{A - A'}{4},
\end{eqnarray}
where $A$ and $A'$ are the areas of the horizon with mass $M$ and mass $M'$, respectively \cite{Chen:2015ibc}.
\end{itemize}

Therefore, the transition probability from a black hole with mass $M$ to mass $M'$ (with emission) is
\begin{eqnarray}
p_{j} \simeq e^{- 4\pi \left( M^{2} - M'^{2} \right)}.
\end{eqnarray}
If $M - M' = \delta M \ll M$, then this probability approximately becomes the Boltzmann factor $e^{- \delta M / T}$, where $T \equiv 1/8\pi M$ recovers the correct Hawking temperature.

Therefore, in some sense, we can conclude that \textit{Hawking radiation is a kind of instantons} \cite{Chen:2018aij}. The Euclidean path integral with a matter field very naturally contains the evaporation process of a black hole, although we need to carefully integrate individual Hawking radiation emission processes successfully.

\subsection{Paths toward trivial geometries}

One further remark is that, in the discussion of the previous subsection, we do not need to assume small $\delta M$. Even with $\delta M$ in the same order of $M$, the computations will be the same as long as we rely on the free scalar field. Therefore, in the extreme limit $M' = 0$, the black hole can transit to a Minkowski background, and the transition probability is
\begin{eqnarray}
p_{j} \simeq e^{- 4\pi M^{2}}.
\end{eqnarray}
This probability is exponentially suppressed, but it is fair to say that this is non-vanishing.

\begin{figure}[b]
\sidecaption
\includegraphics[scale=.55]{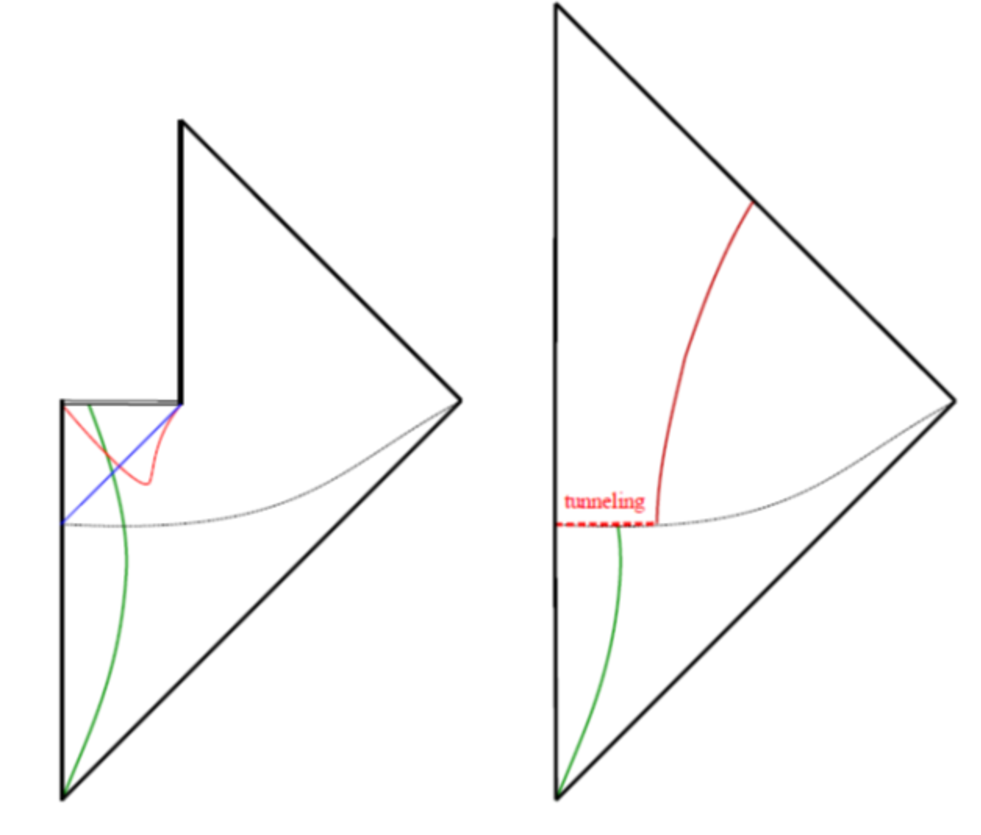}
\caption{Left: The causal structure of a semi-classical spacetime, where the blue line is the event horizon and a thin red curve is the apparent horizon. Right: The causal structure after the tunneling toward the trivial geometry, where we cut the slice (black dotted curve) of the tunneling outside the horizon. The green curve denotes the collapsing matter, and the red curve denotes the emitted matter after the tunneling.}
\label{fig:fig7}
\end{figure}

In terms of the Euclidean manifold, black holes have the topology $D^{2} \times S^{2}$, where $S^{2}$ is due to the spherical symmetry and $D^{2}$ covers radial and (Euclidean) time directions \cite{Hawking:2005kf}. On the other hand, a Minkowski metric has the topology $D^{3} \times S$, where $S$ is the Euclidean time direction that can be identified with an arbitrary period. This periodically identified Euclidean Minkowski is called by a \textit{trivial geometry} because it has no horizon \cite{Hawking:2005kf} (Fig.~\ref{fig:fig7}).

One critical remark is that, as long as we assume a free scalar field, the Euclidean path integral includes a non-vanishing transition channel toward a trivial geometry (hence, results in a topology change). Of course, we needed to assume a free scalar field, but this is not a strong assumption; for example, the same computations can be easily generalized to graviton modes. Therefore, a tunneling channel toward the trivial geometry is a generic consequence of the Euclidean path integral approach.

\subsection{Information survived: importance of trivial geometries}

Then, what is the importance of the trivial geometries? Intuitively speaking, after the Wick-rotation, for non-trivial geometries, there exists a horizon and a singularity; on the other hand, for trivial geometries, there is no horizon nor singularity. Therefore, with trivial geometries, there is no distinction between inside and outside the horizon; also, there is no way to lose information.

This can be demonstrated by computing correlation functions, especially relying on the AdS/CFT correspondence \cite{Maldacena:2001kr}. (All our computations can easily be extended to the anti-de Sitter background.) In an Euclidean black hole background, when we compute correlation functions (say, $\langle \phi \phi \rangle$) as a function of time, this exponentially decays to zero, which might be related to the loss of information \cite{Maldacena:2001kr}. On the other hand, in the periodically identified geometry backgrounds, this correlation function does not vanish, although the probability is exponentially suppressed \cite{Maldacena:2001kr,Hawking:2005kf}. Conceptually,
\begin{eqnarray}
\langle \phi \phi \rangle \simeq p_{1} \langle \phi \phi \rangle_{1} + p_{2} \langle \phi \phi \rangle_{2} + ...,
\end{eqnarray}
where $p_{1}$ is the probability of a non-trivial geometry, $\langle \phi \phi \rangle_{1}$ is the correlation function evaluated at the non-trivial geometry background, $p_{2}$ is the probability to a trivial geometry, and $\langle \phi \phi \rangle_{2}$ is the correlation function evaluated at the trivial geometry background \cite{Sasaki:2014spa,Chen:2015lbp}. Although $p_{1} \sim 1$ and $p_{2} \sim e^{-M^{2}} \ll 1$, as time goes on, $\langle \phi \phi \rangle_{1} \rightarrow 0$, while $\langle \phi \phi \rangle_{2} \rightarrow \mathrm{const.}$, and hence, information will be preserved by the wave function through the trivial geometry.

\subsection{`New' wave-particle duality}

Then, what about the geometries? In this picture, we have two observers. In terms of the unitary observer's point of view, the effective metric is obtained by summing over histories \cite{Sasaki:2014spa,Chen:2015lbp}:
\begin{eqnarray}
\langle g_{\mu\nu} \rangle \simeq p_{1} g_{\mu\nu}^{(1)} + p_{2} g_{\mu\nu}^{(2)} + ...,
\end{eqnarray}
where $g_{\mu\nu}^{(1)}$ is a non-trivial geometry and $g_{\mu\nu}^{(2)}$ is a trivial geometry, where both satisfies the semi-classical equations of motion.

Here, since $p_{1}$ is dominated, the non-trivial geometry is dominated, and hence, the semi-classical description is approximately good. However, the actual geometry $\langle g_{\mu\nu} \rangle$ of a unitary observer cannot satisfy semi-classical equations of motion. Therefore, in terms of the unitary observer, Assumption 1 is satisfied, but Assumptions 2 and 3 are not satisfied. On the other hand, for a dominant semi-classical observer, $g_{\mu\nu}^{(1)}$ satisfies Assumptions 2 and 3, while Assumption 1 is not satisfied.

In conclusion, no observer can see the inconsistency of all assumptions and hence realizes the fifth way of candidate resolutions. A semi-classical observer (like us) will lose information effectively, but the entire wave function of the universe will keep information in principle. This is in some sense a \textit{new version of the wave-particle duality} in quantum gravity.

\newpage

\section{Recovery of the Page curve}
\label{sec:3}

This picture successively resolves the conceptual tensions between assumptions, but we remain the following questions:
\begin{itemize}
\item[--] 1. We can agree that information remains in some part of the wave function. However, is it \textit{sufficient} to recover information, even if the contribution is exponentially small? Can we see the retrieval process of information?
\item[--] 2. What can we say about Assumption 4? Is the relation between the areal entropy and the Boltzmann entropy always true, or can it be violated at some moment?
\end{itemize}
In this section, we will answer these questions. To see the information recovering process clearly, we need to see the late-time behaviors of $p_{i}$.

\begin{figure}[b]
\sidecaption
\includegraphics[scale=.4]{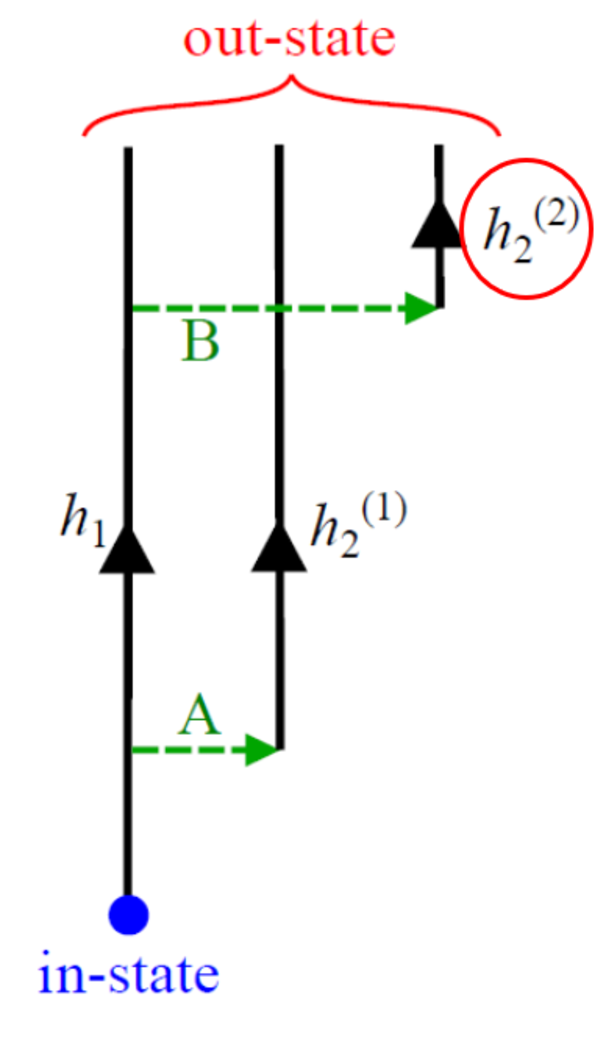}
\caption{A conceptual picture of the quantum gravitational wave function. From the in-state, histories are evolving and branching from bottom to top. $h_{1}$ is the information-losing history, while $h_{2}^{(1)}$ and $h_{2}^{(2)}$ are information-preserving histories, where the tunneling to $h_{2}^{(1)}$ happens in the earlier time and $h_{2}^{(2)}$ happens in the later time. Due to the late-time dominance condition, $h_{2}^{(2)}$ will be dominated eventually and save the information.}
\label{fig:fig8}
\end{figure}

\subsection{Essential conditions for the unitary Page curve}

To see more details of the time evolution of information, it will be sufficient to see the time evolution of the entanglement entropy. Let us define $A$ as being outside the black hole and $B$ as inside the black hole. Initially, there is no Hawking radiation; hence, $S_{A} = 0$. As time passes, the entanglement entropy between the two sides should increase. Eventually, as the black hole evaporation is finished unitarily, there is no interior of the black hole, and hence $S_{A} = 0$, once again. If we can smoothly explain that $S_{A}$ becomes zero, eventually, we can resolve the information loss problem constructively.

Now, to explain this unitary Page curve, we require the following two conditions \cite{Chen:2022ric} (Fig.~\ref{fig:fig8}):
\begin{itemize}
\item[--] 1. \textit{Multi-history condition}: In the computation of entanglement entropies, we need to sum over more than two kinds of contributions (histories, saddles, steepest-descents, or whatever) that dominantly contribute, say $h_{1}$ and $h_{2}$. We further assume that $h_{1}$ is an information-losing history, while $h_{2}$ is an information-preserving history.
\item[--] 2. \textit{Late-time dominance condition}: Initially, the probability of $h_{1}$ (say, $p_{1}$) is dominated by the probability of $h_{2}$ (say, $p_{2}$). However, as time goes on, $p_{2}$ is dominated than $p_{1}$ eventually.
\end{itemize}
If these two conditions are assumed, one can easily explain the unitary Page curve because one can compute the entanglement entropy
\begin{eqnarray}
\langle S \rangle \simeq p_{1} S_{1} + p_{2} S_{2} + ...,
\end{eqnarray}
where in the early time, $p_{2} \ll p_{1} \sim 1$, and in the late time, $p_{1} \ll p_{2} \sim 1$. In the early times, information seems to have disappeared, but eventually, $S_{2}$ goes to zero (because $h_{2}$ is an information-preserving history). Hence, $\langle S \rangle \rightarrow 0$ will be realized.

The critical question is, \textit{does the Euclidean path integral approach realize these two}? The answer is yes.

First, the Euclidean path integral provides multiple histories. For trivial geometries, there is no distinction between inside and outside the horizon, and hence, the entanglement entropy is automatically zero, so to speak, $S_{2} = 0$. For non-trivial geometries, we do not know how to compute entanglement entropies exactly, but let us naively assume that the entanglement entropy monotonically increases: so to speak, $S_{1} = S_{0} - S$, where $S_{0}$ is the initial entropy and $S$ is the areal entropy of the black hole.

The more complicated part is checking the late-time dominance condition. Let us continue to discuss this in the following subsection.

\subsection{Late-time dominance of trivial geometries}

Now, let us compute the probabilities toward trivial geometries. Of course, this is not a trivial process and depends on the detailed tunneling mechanism. However, we can provide a reasonable guideline for computing the time-dependent probabilities even with this limitation.

First, as we have observed in the previous subsection, Hawking radiation can be regarded as a kind of instantons \cite{Chen:2018aij}, and, in some limit, such a free scalar field instanton can describe a tunneling process toward a trivial geometry. Therefore, the existence of such a channel at any time is a generic consequence. At each time, the tunneling probability toward the trivial geometry is $\sim e^{-S}$, where $S$ is the Bekenstein-Hawking entropy at the time.

Second, the Euclidean time period is not unique. Of course, the probability decreases as the Euclidean time period increases; hence, we can ignore it in general. However, as the black hole size decreases, the contribution from the multiple time periods may contribute more and more: the tunneling rate toward a trivial geometry is then
\begin{eqnarray}
\Gamma = \sum_{n=1}^{\infty} e^{-S (2n - 1)} = \frac{1}{e^{S} - e^{-S}},
\end{eqnarray}
where $n$ is introduced to present the multipleness of the Euclidean time period \cite{Chen:2022ric}.

Finally, let us assume that as we average over all other non-trivial geometries, we obtain the semi-classical history, which is an information-losing history. The sum of all histories must be a unity, and the tunneling rate from a semi-classical history to a trivial geometry is $\Gamma$. If the probability of the semi-classical history is $p_{1}$ and the probability toward a trivial geometry is $p_{2}$, then $p_{1} + p_{2} = 1$ and $\Gamma = p_{2}/p_{1}$. From this, we obtain
\begin{eqnarray}
p_{1} &=& \frac{e^{S} - e^{-S}}{1 + e^{S} - e^{-S}},\\
p_{2} &=& \frac{1}{1 + e^{S} - e^{-S}}.
\end{eqnarray}

\begin{figure}[b]
\sidecaption
\includegraphics[scale=.27]{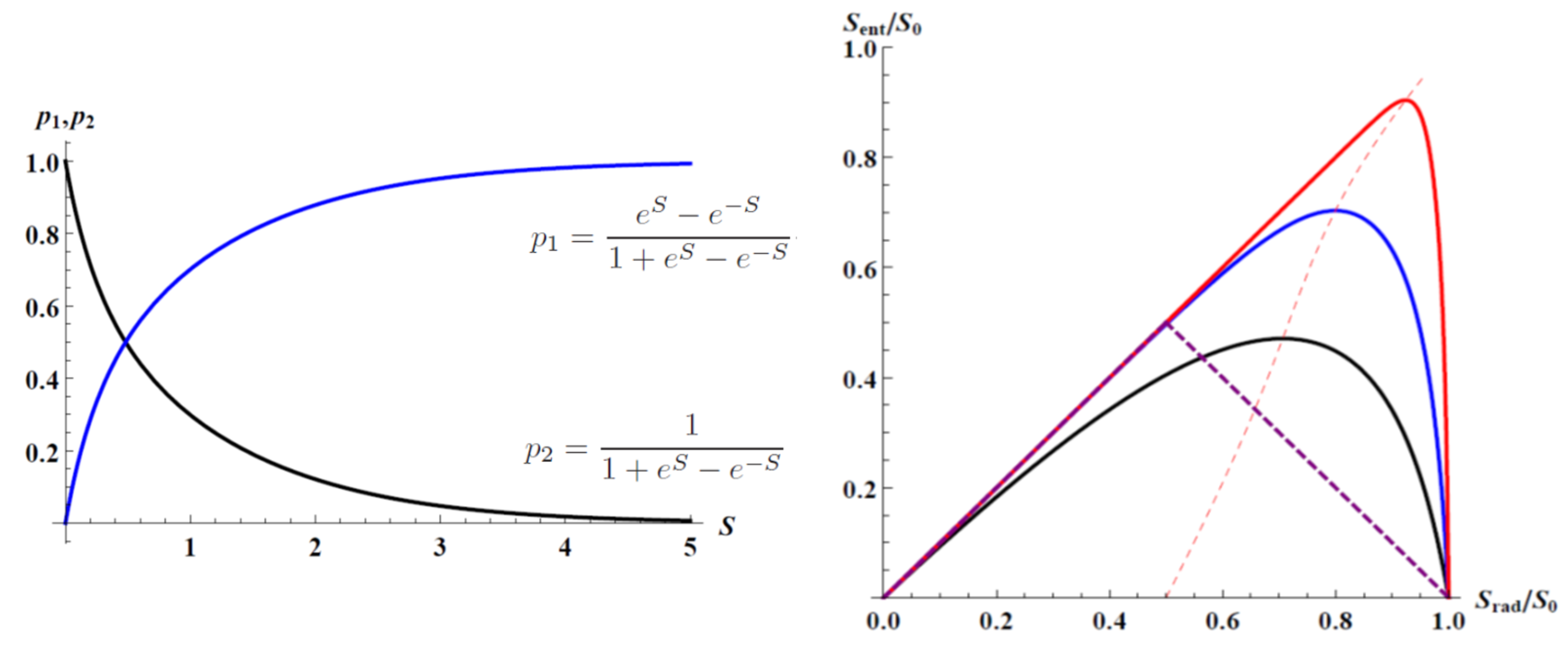}
\caption{Left: $p_{1}$ and $p_{2}$ as a function of $S$. The blue and black curves cross around the Planck scale. Right: $\langle S \rangle$, where black, blue, and red curves correspond larger and larger $S_{0}$ \cite{Chen:2022eim,Chen:2022ric}.}
\label{fig:fig9}
\end{figure}

These probabilities were computed with many simplifications but still contained our expected essential features. In the beginning, i.e., if $S \gg 1$, $p_{1}$ is much more dominated, and $p_{2}$ is exponentially suppressed. However, in the late time, as $S$ approaches to the Planck scale, the golden cross between $p_{1}$ and $p_{2}$ appears, and eventually, in the $S \rightarrow 0$ limit, $p_{1} \rightarrow 0$ while $p_{2} \rightarrow 1$. This realizes the \textit{late-time dominance} from the Euclidean path integral approach (left of Fig.~\ref{fig:fig9}).

\subsection{Recovery of the Page curve}

As a result, we can compute the entanglement entropy as follows \cite{Chen:2022eim,Chen:2022ric}:
\begin{eqnarray}
\langle S \rangle &=& p_{1} S_{1} + p_{2} S_{2}\\
&=& \left( S_{0} - S \right) \left( \frac{e^{S} - e^{-S}}{1 + e^{S} - e^{-S}} \right),
\end{eqnarray}
where we assumed that $S_{0}$ is the initial Bekenstein-Hawking entropy, $S_{1} = S_{0} - S$ is the radiation entropy, and hence, the entanglement entropy of the semi-classical history monotonically increases, and $S_{2} = 0$, because there is no interior of the black hole in the trivial geometry. This explains the unitary Page curve (right of Fig.~\ref{fig:fig9}).

However, there are some important remarks in this final result.
\begin{itemize}
\item[--] First, there exists a moment when the entanglement entropy goes beyond the entropy bound. In some sense, Assumption 4 is violated instantly because the entanglement entropy might be greater than the areal entropy. However, this does not cause the infinite production problem because the entropy accumulation is just the order of the initial entropy. Also, for a given semi-classical history, the unitarity is effectively broken, and hence, it is not surprising that there is no apparent problem even with the violation of Assumption 4.
\item[--] Second, we assumed $S_{1} = S_{0} - S$, but this is not necessarily true in general. This might overestimate the entanglement entropy of the semi-classical history. If a new mechanism can take entanglements from inside to outside or breaks maximum entanglements, it is not surprising that $S_{1}$ might be modified; this will be partly helpful to recover the entropy-bound relation.
\item[--] Third, if multiple channels exist toward the trivial geometry \cite{Chen:2023voq}, the transition to tunnel to the trivial geometry becomes higher. Hence, it will be helpful to decrease the maximum value of the entanglement entropy.
\end{itemize}

In conclusion, our approach can explain the unitary Page curve within very reasonable and conservative assumptions of quantum gravity. This nicely reveals the analog relations between the wave and particle duality in the quantum gravitational version, where this is the crucial idea to overcome inconsistencies between assumptions. There exists a moment when the entropy bound is violated (hence, Assumption 4 is violated), but it seems that this has not caused any serious problems yet.

Can the Euclidean path integral \textit{does} solve the information loss problem? Do we still have any uncertain issues? In the next section, we will finally summarize future perspectives.

\section{Future perspectives}
\label{sec:4}

\subsection{Comparison with alternative ideas}

First, let us compare this Euclidean path integral picture to various candidate ideas to resolve the information loss problem.

\subsubsection{Black hole complementarity and black hole firewalls}

The black hole complementarity principle was proposed initially to overcome the inconsistency between Assumptions 1 - 5 \cite{Susskind:1993if}. If there is no witness to experience inconsistencies between assumptions, it is still consistently acceptable; this is somehow related to the complementarity principle of wave and particle natures. However, the problem is that it is possible to construct an observer who experiences inconsistency between assumptions, as discussed in the previous section \cite{Yeom:2008qw,Hong:2008mw,Yeom:2009zp}.

However, the original intuition might be correct in some sense. Like the wave and particle natures, the assumptions are intuitively inconsistent. However, if a \textit{measurement} divides wave and particle nature, one can avoid the apparent inconsistency between contradictory properties. The original black hole complementarity principle was wrong because everything was drawn in the same causal diagram. On the other hand, in our wave function picture, there are two points of view: one is a unitary observer who sees the entire wave function which resembles a wave nature, and the other is a semi-classical observer who sees only a single semi-classical history, that resembles with a particle nature \cite{Sasaki:2014spa,Chen:2015lbp}. \textit{Two observers are not living in the same causal diagram} \cite{Hartle:2015bna}; hence, it is impossible to construct an inconsistency experiment, which was possible in the original black hole complementarity.

As a result, if we think about the unitary observer, who experiences the unitarity of the entire process, the observer will notice that the semi-classical equations are not satisfied; unitarity enforces the violation of general relativity even in the horizon scale. This might be related to the \textit{black hole firewall} \cite{Almheiri:2012rt}. Initially, the effects of the firewall were regarded to be restricted inside the horizon, but this is not logically natural. In our Euclidean wave function picture, it is not surprising, even though effects from the firewall reach infinity \cite{Hwang:2012nn,Kim:2013fv,Chen:2015gux}.

\subsubsection{Quantum extremal surfaces and replica wormholes}

A proposal was made that the entanglement entropy can be computed using the replica trick \cite{Almheiri:2019qdq}. Before we compute the entanglement entropy, we first compute the Renyi entropy, and a specific limit of the Renyi entropy corresponds to the entanglement entropy. It was interesting to observe that there are two main contributions to the Renyi entropy; one is the Gibbons-Hawking-like contribution that explains the increasing part of the entanglement entropy, and the other is a new type of contribution, so-called \textit{replica wormholes} \cite{Almheiri:2020cfm}.

Then, which contribution is dominant when we compute the entanglement entropy? In terms of \textit{quantum extremal surfaces}, there is only one contribution before the Page time; hence, the entanglement entropy monotonically increases. However, two quantum extremal surfaces appear after the Page time, and we need to choose only one to compute the entanglement entropy. Eventually, one can explain the Page curve that we expected \cite{Almheiri:2019hni}.

However, there are still some uncertain questions about this approach. For example, the new contributions, so-called replica wormholes, are only observed in terms of the density matrix \cite{Penington:2019kki}; that means there is no corresponding contribution in the wave function itself, although the wave function is a more fundamental notion than the density matrix. Interestingly, for some examples with a holographic dual picture, one can see that the corresponding quantum extremal surfaces before and after the Page time are changed in its dual pictures \cite{Almheiri:2020cfm}; however, there is no smooth time-dependent description between the two limits before and after the Page time. Perhaps, to describe such a smooth time evolution around the Page time, we need to know the wave function rather than the density matrix.

However, it is fair to say that this picture shares the same philosophy as the Euclidean path integral picture \cite{Chen:2022eim,Chen:2022ric}. We remark on the two essential conditions. First, we need a multi-history condition; in the replica trick, there are more than two types of contribution, one information-losing (Gibbons-Hawking contributions) and the other information-preserving (replica wormholes). Second, we need a late-time dominance condition; after the Page time, we only select the contributions from the information-preserving history. Therefore, we share the same idea to resolve the information loss paradox: information will be preserved by the non-perturbative contributions of the entire path integral. However, there is a difference in whether the path integral is of the wave function (our picture) or the density matrix (replica trick).

\begin{figure}[t]
\sidecaption
\includegraphics[scale=.5]{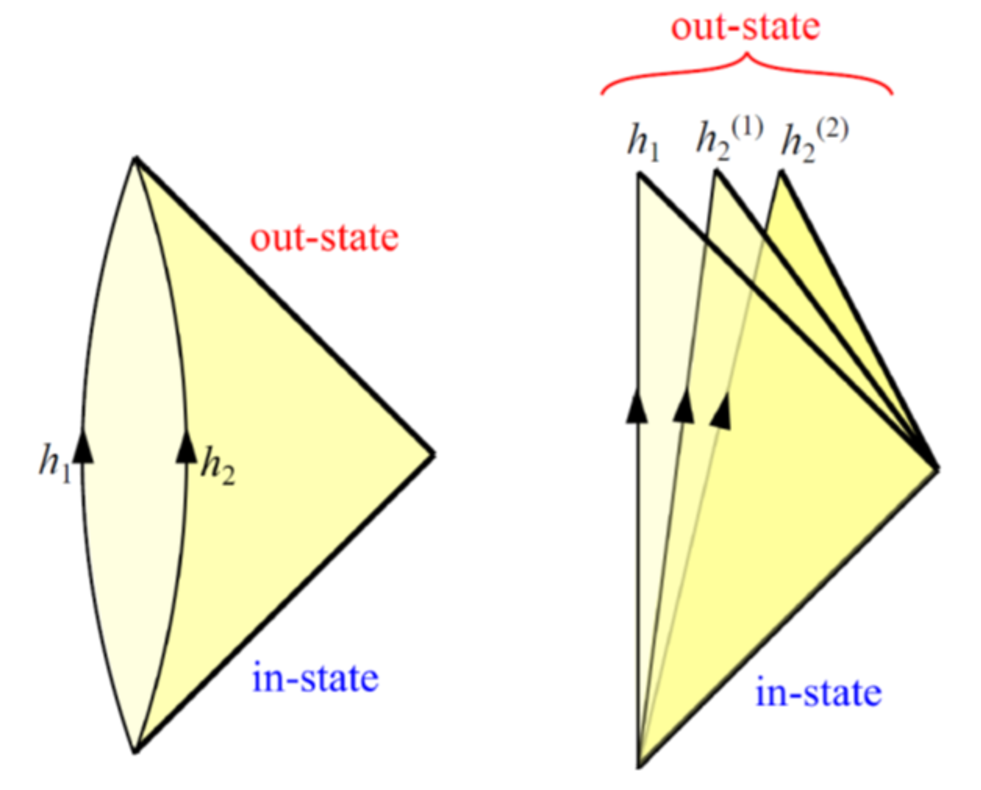}
\caption{Left: We sum over all histories by fixing the boundary. Right: Without fixing the boundary, we follow the bulk history branching process.}
\label{fig:fig10}
\end{figure}

\subsubsection{Bulk-to-boundary vs. boundary-to-bulk}

The most substantial evidence of the unitarity of black hole evolution is the AdS/CFT correspondence or the holographic principle \cite{Maldacena:1997re}. The physics of bulk and boundary correspond, while all physical boundary descriptions are manifestly unitary. Therefore, it is evident that the bulk dynamics, including the formation and evaporation of a black hole, are also unitary, even though there is no constructive description yet.

However, it is important to note one interesting philosophical difference between the typical holography and quantum cosmology: \textit{whether we fix the boundary first or the bulk first} (Fig.~\ref{fig:fig10}). In the usual approaches of holography, we first fix the boundary and sum over all bulk contributions that satisfy the fixed classical boundary. On the other hand, in the usual quantum cosmological approaches, diverse boundaries can be branched from a given quantum state \cite{Hartle:2015bna}, which is known as the \textit{Everett branching} \cite{Everett:1957hd}. Hence, the final boundary is eventually a superposition of various classical boundaries \cite{Sasaki:2014spa,Chen:2015lbp}.

For example, in the Euclidean path integral approach, considering bubble nucleation from the bulk side may eventually change the boundary \cite{Chen:2016nvj}. However, if we first fix the boundary at infinity, we must neglect such a process. So, the ER=EPR conjecture might be falsified for the former case \cite{Chen:2016nvj,Chen:2020nag}, while in the latter case, the ER=EPR conjecture cannot be violated \cite{Maldacena:2013xja}.

Then, we can ask which approach is more suitable for unitary time evolution. Perhaps both approaches should have consistent unitary descriptions, and it will be worthwhile to study each approach independently. We leave this for an interesting future research topic.

\subsubsection{Remnant picture and entropy bounds}

We noticed that our approach violates the entropy bound for a while, i.e., the statistical entropy is larger than its Bekenstein-Hawking entropy. Is this possible in principle? Interestingly, there is no fundamental proof to secure this entropy-bound relation. Our approach is not the same as the Planck-scale remnant scenario \cite{Chen:2014jwq} but shares one point: to allow the violation of the entropy bound.

Within the semi-classical framework, it is possible to construct a thought experiment or an algorithm that explicitly violates the entropy bound \cite{Bae:2020lql,Buoninfante:2021ijy}. However, this does not seem natural from a holographic point of view. Also, several observations show that the Bekenstein-Hawking entropy correctly provides the Boltzmann entropy. However, these computations rely on a zero-temperature system \cite{Strominger:1996sh}, while we need fully dynamic situations. It is fair to say that the belief in the entropy-bound relation cannot be strongly supported in dynamic situations. Can we have a consistent point of view between the semi-classical and holographic approaches? We also leave this question for future research.

\subsection{Conclusion: is it sufficient?}

This article introduced the wave function using the Euclidean path integral. First, we want to emphasize that the wave function picture can explain the tensions among the assumptions of unitary quantum mechanics and general relativity. The unitarity is about the wave nature, while the semi-classical physics is about the particle nature. We cannot observe them at the same time. Hence, a unitary observer collects all histories and loses the semi-classicality. In contrast, a semi-classical observer like us will satisfy local quantum field theory and general relativity while effectively losing unitarity.

We can explain the information loss problem more technically. A summation of instantons can approximate the Euclidean path integral. The instantons include not only Hawking radiation but also a non-perturbative process that explains tunneling from a black hole to a Minkowski or trivial geometry. If there is a black hole, it is unsurprising to lose information, and the entanglement entropy will increase monotonically. On the other hand, if the spacetime tunnels to a trivial geometry, there is no horizon; hence, there is no way to lose information, and the entanglement entropy will be zero. We have shown that the existence of such a tunneling channel is evident within very natural assumptions.

Moreover, we computed the time dependence of each history, where one is information-losing, and the other is information-preserving. We observed that initially, the former is dominated, while finally, the latter is dominated. Therefore, we can recover a unitary Page curve, although the entropy-bound relation seems to be violated for a while.

Finally, we must mention that we relied on the Euclidean analytic continuation, although the trivial geometry preserves information; hence, we only considered the propagation of geometry outside the event horizon. Using the Lorentzian time description, can we explicitly extend a consistent description that covers both inside and outside the horizon? This may require a method to solve the Wheeler-DeWitt equation \textit{directly} rather than using the path integral. Explicitly, we mention possible future research topics.
\begin{itemize}
\item[--] Can we explain the same process by directly solving the Wheeler-DeWitt equation without relying on the Euclidean path integral?
\item[--] In this Lorentzian time consideration, can we explain the time evolution of slices that cover both inside and outside the horizon?
\item[--] Can we explain information propagation more constructively?
\item[--] Can we constructively explain the late-time dominance condition independently and consistently in this direct approach?
\item[--] In this case, what is the role or importance of the singularity? Should we provide any boundary condition at the singularity, e.g., the DeWitt boundary condition \cite{Bouhmadi-Lopez:2019kkt,Brahma:2021xjy}, or does this have nothing to do with the unitary time evolution?
\item[--] What is the relationship with various ideas that resolve the singularity, e.g., regular black hole pictures \cite{Bouhmadi-Lopez:2020wve} or loop quantum gravity pictures \cite{Bojowald:2018xxu,Brahma:2020eos}? Can we provide a consistent picture that incorporates the Euclidean path integral approach?
\end{itemize}
We will reach the ultimate and consistent answer to the information loss paradox as we answer these questions.

\newpage

\begin{acknowledgement}
This work is an invited chapter for the edited book ``The Black Hole Information Paradox'' (Eds. Ali Akil and Cosimo Bambi, Springer Singapore, expected in 2025). This work was supported by the National Research Foundation of Korea (Grant No. : 2021R1C1C1008622, 2021R1A4A5031460).
\end{acknowledgement}

\end{document}